\documentclass[article,aps,pra,twocolumn,groupedaddress,showpacs]{revtex4}
\usepackage{amsmath}
\usepackage{amsfonts}
\usepackage{amssymb}
\usepackage{graphicx}%

\begin{document}

\title{Fluctuations assisted stationary entanglement in driven quantum systems} 

\author{Dimitris G. Angelakis}
\email{dimitris.angelakis@gmail.com}
\address{
Science Department, Technical University of Crete, 73100 Chania, Greece EU
\\
Centre for Quantum Technologies, National University of
Singapore,  117543 Singapore}

\author{Stefano Mancini}
\email{smancini66@gmail.com}

\address{School of Science and Technology, University of Camerino, 62032 Camerino, Italy EU}

\pacs{03.67.Bg, 03.65.Yz, 42.50.Lc}


\begin{abstract}
We analyze the possible quantum correlations between two coupled dimer systems  
 in the presence of independent losses and driven by a fluctuating field.  
For the case of the interaction being of a Heisenberg exchange type, we first
 analytically show the possibility for robust stationary entanglement for realistic values of the dissipation rates
 and then analyze its robustness as a function of the noise to signal ratio of the pump.
 We find that for a common fluctuating driving field,  a stochastic resonance effect  appears as function of the ratio between
 field strength and noise strength. The effect disappears in the case of
 uncorrelated or separate pumps.
Our result is general and could be applied to different quantum systems ranging from electron spins in solid state, to ions trap technologies and cold 
atom set ups.
	
\end{abstract}


\maketitle

{\it Introduction:}  Entanglement is nowadays commonly considered a resource for quantum information processing \cite{vedral07}. Being of a purely quantum  nature can be easily degraded by reservoir contaminations.  
Nevertheless, recently the possibility of having stationary entanglement in open quantum systems has been put forward \cite{braun02,PH02, clark03,mem11}, also achieving experimental evidence \cite{polzik10}.
However it is not clear how robust is such stationary entanglement against other sources of noise that can eventually affect the system, especially when originating from the pump.
 
In this work we aim at clarifying the robustness of stationary  entanglement in lossy coupled systems in the presence of noise in the amplitude fluctuations of the driving field. We assume the case of the always on exchange interactions between the systems,  usually employed when dealing with local dissipative environments\cite{clark03}.
The entanglement robustness is characterized by means of a figure of merit related to the quantum concurrence and accounts for the signal to noise ratio of the driving field. We distinguish the case of local and global fluctuations and show that in the latter case stochastic resonance like effects \cite{march} can emerge for an optimal ratio between the strength and the noise
of the common driving field. The effect disappears in the case of uncorrelated or separate pumps.

Our analysis is general, the results are analytic, and could be applied to different set ups ranging from electron spins in double quantum dots  or  in effective spin models generated in driven ions, cold atoms and  coupled cavity arrays ups\cite{dots,ions,coldatoms,CCAs}


{\it The system: } We assume two coupled dimer systems labelled by $A$ and $B$  interacting through an exchange interaction. The 
systems are subject to incoherent driving of amplitude $\alpha$ from outside and dissipate independently into two separate reservoirs 
with equal dissipation rates given by $\gamma$(see Fig. 1). The Hamiltonian describing the coherent part of the interaction in this case is:
\begin{equation}
H=J(\sigma_A^{\dag}\sigma_B+\sigma_A\sigma_B^{\dag})+\alpha (\sigma_A^y+\sigma_B^y),
\label{Hp}
\end{equation}
with $\sigma$s the usual Pauli operators on $\mathbb{C}^2$ such that $\sigma=(\sigma^{x}+\iota\sigma^{y})/2$ and $2\sigma^{\dag}\sigma-I=\sigma^z$.

The open system dynamics will be described by the master equation 
\begin{eqnarray}
\dot\rho=-\iota [H,\rho]
+\mathcal{D}{[\sigma_A]}\rho+\mathcal{D}[\sigma_B]\rho,
\label{me1}
\end{eqnarray}
where
\begin{equation}
\mathcal{D}[a]b:=\gamma(2aba^{\dag}-a^{\dag}ab-ba^{\dag}a),
\end{equation}
denotes the dissipative super-operator and $\gamma$ is the decay rate into the two separate environments.

In this work besides looking for entanglement at steady state, we would also investigate its robustness against fluctuations in the driving field.
We thus assume that the driving field on top of its coherent amplitude $\alpha$, exhibits a Gaussian white noise term 
$\xi(t)$, with
\begin{equation}
\label{white}
\langle\xi(t)\rangle=0,\quad\langle\xi(t)\xi(t^{\prime})\rangle=2\eta\delta(t-t^{\prime}),
\end{equation}
and $\eta\ge0$ measuring the noise strength. 
\begin{figure}
\centering
\includegraphics[width=0.4\textwidth]{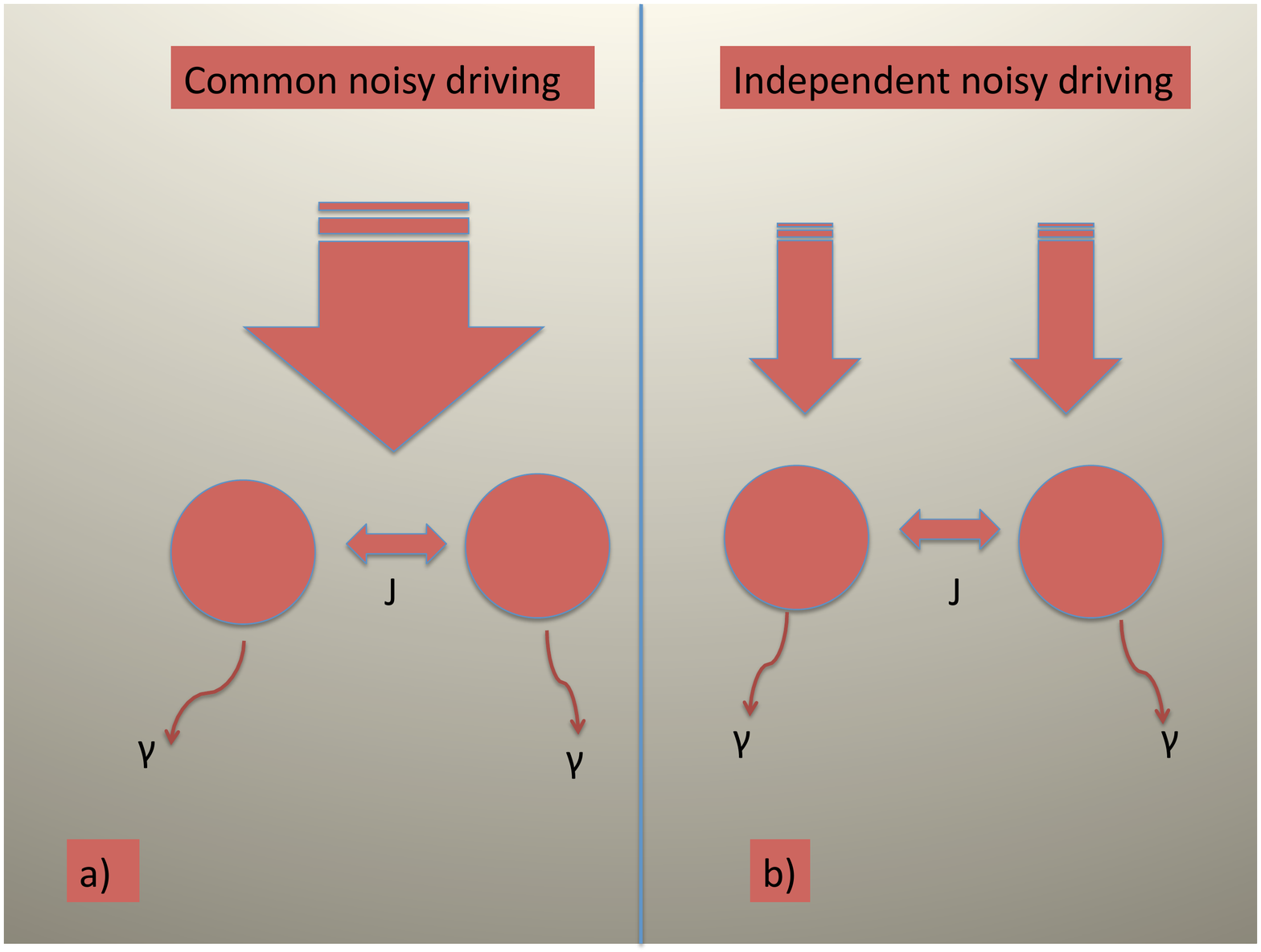}
\caption{The system under consideration. Two coupled dimer systems (atoms or quantum dots or spins) are interacting and dissipating in different reservoirs. 
In b) two uncorrelated noisy pumps are continuously driving the systems whereas
in a) the same pump is driving both.   } \label{system}
\end{figure}

Hence the master equation \eqref{me1} will become 
\begin{eqnarray}
\dot\rho=-\iota [H,\rho]
+\mathcal{D}[\sigma_A]\rho+\mathcal{D}[\sigma_B]\rho+\mathcal{N}[\sigma_A^y;\sigma_B^y]\rho,
\label{me2}
\end{eqnarray}
with $\mathcal{N}$ the superoperator describing the noisy effects of $\xi(t)$.

The steady state solution of Eq.(\ref{me2}) can be found by
writing the density operator and the other operators in a matrix
form, in the basis
$\mathbb{B}=\{|0\rangle_{A}|0\rangle_{B},|0\rangle_{A}|1\rangle_{B},
|1\rangle_{A}|0\rangle_{B},|1\rangle_{A}|1\rangle_{B}
\}$. Let us parametrize the density operator as
\begin{equation}
\rho_{ss}= \left(\begin{array}{cccc} a&b_{1}+\iota b_{2}&
c_{1}+\iota c_{2}&d_{1}+\iota d_{2}
\\
b_{1}-\iota b_{2}&e& f_{1}+\iota f_{2}&g_{1}+\iota g_{2}
\\
c_{1}-\iota c_{2}&f_{1}-\iota f_{2}& h&i_{1}+\iota i_{2}
\\
d_{1}-\iota d_{2}&g_{1}-\iota g_{2}& i_{1}-\iota i_{2}&1-a-e-h
\end{array}\right).
\end{equation}

Once we know the steady state density matrix,
to quantify the entanglement we will use the concurrence \cite{Woo}
\begin{equation}
C(\rho_{ss}):=\max\left\{0,\lambda_1-\lambda_2-\lambda_3-\lambda_4\right\},
\end{equation}
where $\lambda_i$'s are, in decreasing order, the nonnegative square
roots of the moduli of the eigenvalues of $\rho_{ss}\tilde\rho_{ss}$ with
\begin{equation}
\tilde\rho_{ss}:=\left(\sigma_{1}^{y}\otimes \sigma_{2}^{y}\right)\rho_{ss}^*\left(\sigma_{1}^{y}\otimes \sigma_{2}^{y}\right),
\end{equation}
and $\rho_{ss}^*$ denotes the complex conjugate of $\rho_{ss}$.


We will separate our analysis in two cases, the common driving case where both systems are subject to the same fluctuating pump and the case
when independent driving is applied to each one.

{\it Common driving:} If the driving field is common to both systems (as also in  \cite{ALK10}), then
the two systems experience the same fluctuations.
Hence the superoperator $\mathcal{N}$ reads (see also \cite{GK76})
\begin{eqnarray}
\mathcal{N}[\sigma_A^y;\sigma_B^y]\rho=
-\eta\left[(\sigma_A^y+\sigma_B^y),\left[(\sigma_A^y+\sigma_B^y),\rho\right]\right].
\end{eqnarray}
We can then analytically solve the linear system of equations coming from the master equation \eqref{me2}
\begin{eqnarray}
\label{leq1}
0=-\iota J[(\sigma_A^{\dag}\sigma_B+\sigma_A\sigma_B^{\dag}),\rho]
-\iota\alpha\left[(\sigma_A^y+\sigma_B^y),\rho\right] \nonumber \\
-\eta\left[(\sigma_A^y+\sigma_B^y),\left[(\sigma_A^y+\sigma_B^y),\rho\right]\right]
+\mathcal{D}[\sigma_A]\rho+\mathcal{D}[\sigma_B]\rho,
\end{eqnarray}
for $a,b_1,b_2,c_1,c_2,d_1,d_2,e,f_1,f_2,g_1,g_2,h,i_1,i_2$.
The explicit solution is reported in Appendix A.
In Figure \ref{concur1} we plot the concurrence $C$ \cite{Woo} vs amplitude of the pump  $\alpha$ and 
fluctuation $\eta$ for a fixed nonzero values of $J$.  

\begin{figure}
\centering
\includegraphics[width=0.4\textwidth]{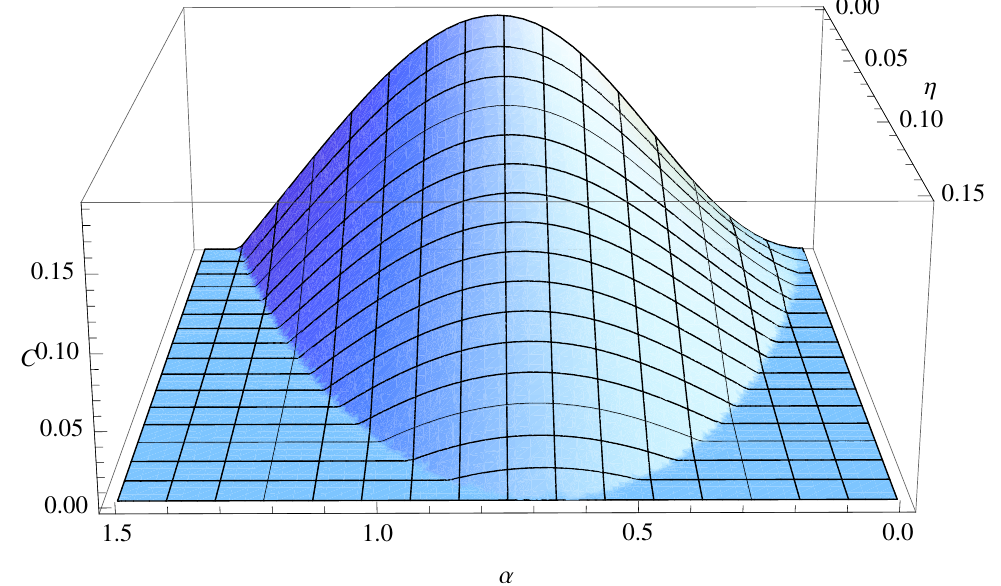}
\caption{Steady state entanglement as quantified by concurrence $C(\alpha,\eta)$ vs $\alpha$ and $\eta$ for $J=2$  in units of the dissipation rate $\gamma$.} \label{concur1}
\end{figure}

\begin{figure}
\centering
\includegraphics[width=0.4\textwidth]{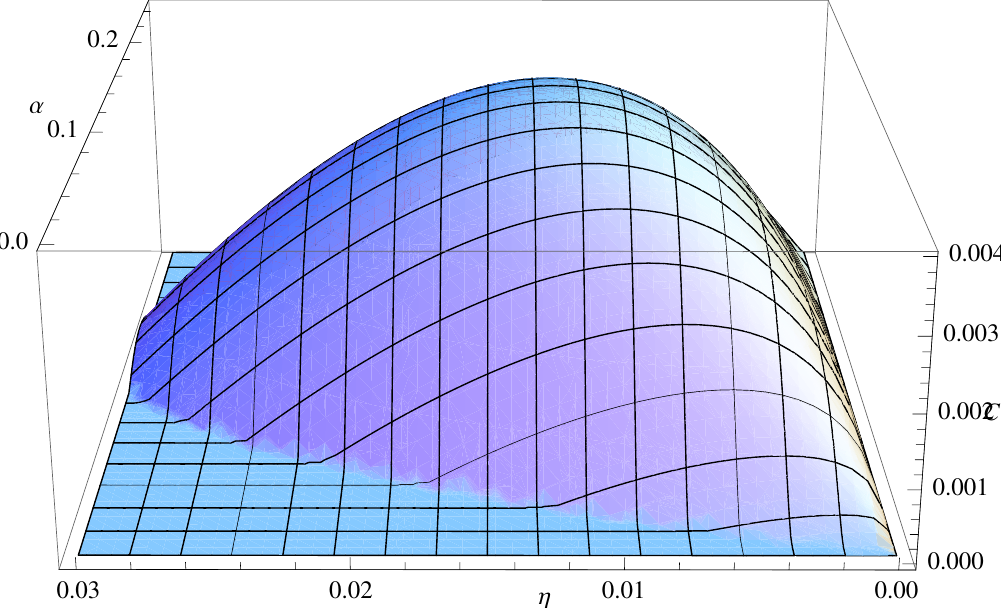}
\caption{The quantity $\Delta(\alpha,\eta)$ vs $\alpha$ and $\eta$ for $J=2$ (units  as in Fig. 2). 
} \label{concur_diff1}
\end{figure}

We note here that $C$ is not always monotonically decreasing vs $\eta$ which can been seen by introducing 
 the quantity
\begin{equation}
\label{Delta}
\Delta(\alpha,\eta)=
C(\alpha,\eta)-C(\alpha,\eta=0)
\end{equation}
for $C(\alpha,\eta)-C(\alpha,\eta=0)>0$,
which quantifies the difference (if positive) between the concurrence in presence and absence of fluctuations in the driving field. $\Delta(\alpha,\eta)=0$ when $C(\alpha,\eta)-C(\alpha,\eta=0)\le0$
The inverse U shape of such a quantity vs $\eta$, as shown in Fig. \ref{concur_diff1}, indicates a stochastic resonance-like effect \cite{march}. That is, there is an optimal non zero value of noise strength for the driving field which maximize the concurrence.
In Fig. \ref{concur1} we may also observe that entanglement is quite robust in the range of pumping slightly larger than the decay rate and it survives for
white noise up to 10$\%$ in the fluctuating field. 

\begin{figure}
\centering
\includegraphics[width=0.4\textwidth]{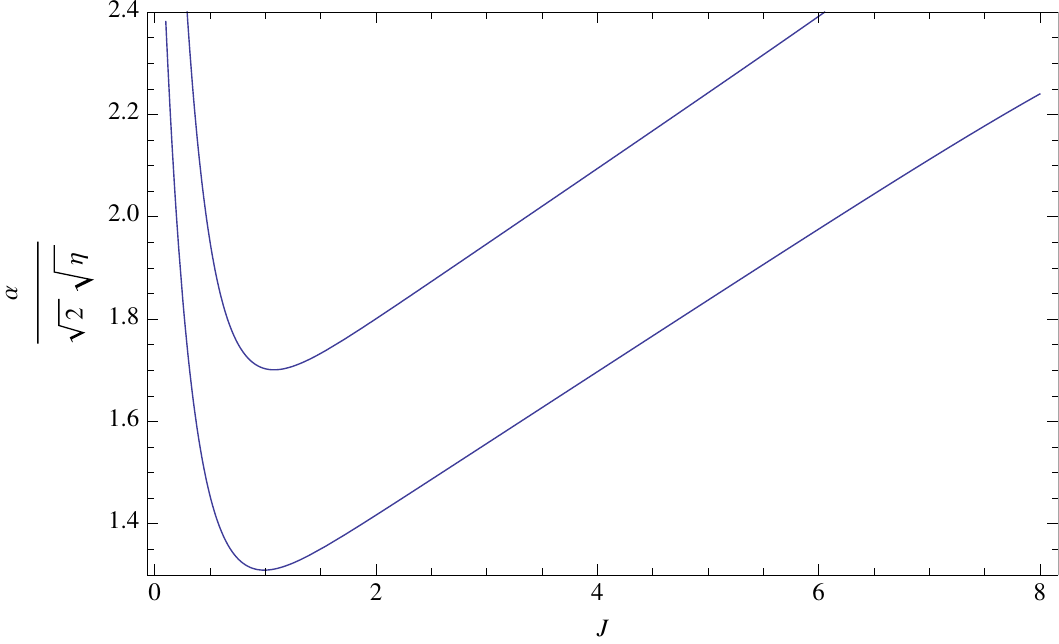}
\caption{
The quantity $SNR$ is plotted vs $J$  in units of the dissipation rate $\gamma$.
Top (bottom) curve refers to the separable (non separable) case.} \label{signoi}
\end{figure}

To quantify the robustness of entanglement versus noise in the driving field we may consider the signal to noise ratio  
$SNR:=\alpha/\sqrt{2\eta}$ for $\alpha$ corresponding to $\max_{\alpha} C(\alpha,\eta=0)$ and $\eta$ corresponding to  the value for which $\max_{\alpha} C(\alpha,\eta)$ becomes zero. The smaller is this quantity, the more robust is entanglement.
The $SNR$ has been plotted in 
Figure \ref{signoi} (bottom curve) vs $J$. 
The value $J=1$ is optimal in the sense that admits a larger interval of $\eta$ values for which $\max_{\alpha}C(\alpha,\eta)>0$.

{\it Independent driving  fields: }
If the driving field is separate for the two systems, then they will experience independent fluctuations.
Hence the superoperator $\mathcal{N}$ reads (see also \cite{GK76})
\begin{eqnarray}
\mathcal{N}[\sigma_A^y;\sigma_B^y]\rho=
-\eta\left[\sigma_A^y,\left[\sigma_A^y,\rho\right]\right]
-\eta\left[\sigma_B^y,\left[\sigma_B^y,\rho\right]\right].
\end{eqnarray}
We can then again analytically solve the linear system of equations coming from the corresponding master equation for the steady state
\begin{eqnarray}
\label{leq2}
0=-\iota J[(\sigma_A^{\dag}\sigma_B+\sigma_A\sigma_B^{\dag}),\rho]
-\iota\alpha\left[(\sigma_A^y+\sigma_B^y),\rho\right] \nonumber \\
-\eta\left[\sigma_A^y,\left[\sigma_A^y,\rho\right]\right]
-\eta\left[\sigma_B^y,\left[\sigma_B^y,\rho\right]\right]
+\mathcal{D}[\sigma_A]\rho+\mathcal{D}[\sigma_B]\rho,
\end{eqnarray}
for $a,b_1,b_2,c_1,c_2,d_1,d_2,e,f_1,f_2,g_1,g_2,h,i_1,i_2$ (the solution is reported in Appendix B) and fully characterize the state of the system.

\begin{figure}
\centering
\includegraphics[width=0.4\textwidth]{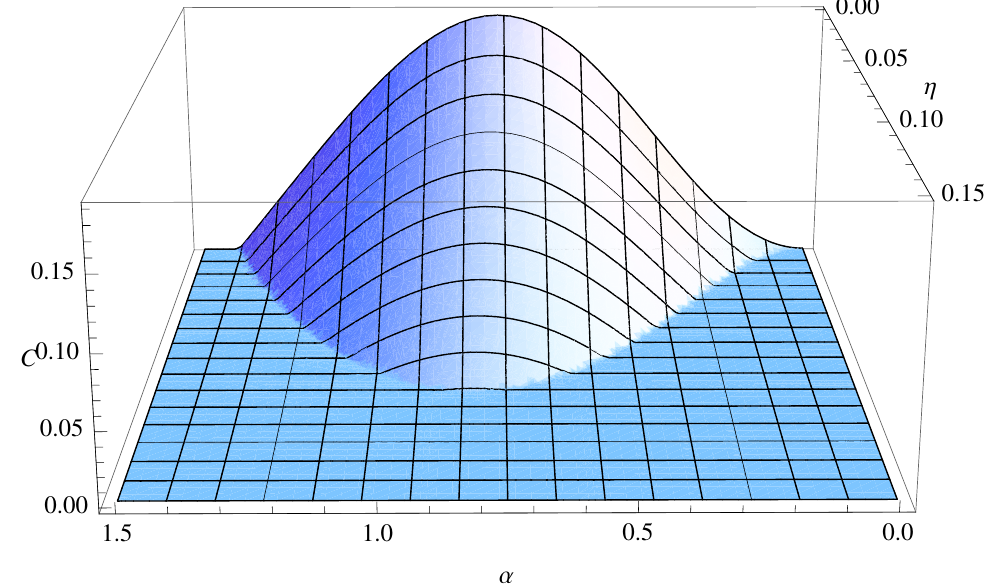}
\caption{Concurrence $C(\alpha,\eta)$ vs $\alpha$ and $\eta$ for $J=2$.} \label{concur2}
\end{figure}

In Figure \ref{concur2} we plot as before the concurrence vs  the pump strenthg $\alpha$ and noise $\eta$ for a fixed nonzero values of $J$. In this case $C$ is a monotonically decreasing function of $\eta$.
Hence, we do not have any stochastic resonance like effects, as matter of fact the quantity \eqref{Delta} is zero everywhere for all values of signal to noise ratio.

By comparing Figure \ref{concur2} and \ref{concur1} we may realize that entanglement is less robust in this case with respect to the previous one. As matter of fact, the maximum value of concurrence decreases to zero
when the signal-to-noise ratio is $\approx 1.8$ in contrast to the previous case where the corresponding ratio was $\approx 1.4$.
In Figure \ref{signoi} (top curve) we have plotted 
the quantity $SNR$ vs $J$. Likewise the previous case the optimal value is $J=1$.

Our results in general are in agreement with similar phenomena that arise in spin chains from the interplay of dissipative and dephasing noise sources \cite{huelga09} and to those introduced in Ref.\cite{PH02}.
In the latter, two dissipating cavity modes interacting with an atom were driven by thermal noise.
However, also in this case the additive noise besides the local dissipative fluctuations is common to both systems,
hence we can quite generally argue that stochastic resonance effects on entanglement arise from the interplay of separate and common fluctuations of different kind. Moreover, we can speculate that non-white noise in Eq.\eqref{white} could lead to more pronounced stochastic resonance like effects and would like to study this in follow ups works.


{\it Conclusion:}
We have studied the robustness of entanglement that can be created in the steady state of a driven two coupled dimers interacting with an exchange interaction.
As a further noise source we have considered amplitude fluctuations of the driving field. 
The entanglement robustness has been characterized by means of a figure of merit accounting for the signal to noise ratio of the driving field. We have distinguished the case of local and global fluctuations and showed that in the latter case stochastic resonance like effects can emerges. Our analysis could be applied to effective exchange interaction models between dimers realizable in various implementations in solid state,  ions traps, cold atoms and Cavity QED set ups.


\begin{widetext}

\appendix

\section{Solution of Eq.(\ref{leq1})}

\begin{eqnarray}
D&=&24 \alpha ^6+J^4 (1+2 \eta )^2 (1+6 \eta )+4 \alpha ^4 \left(15+82 \eta +144 \eta ^2\right)+2 (\alpha +2 \alpha  \eta )^2 \left(21+184 \eta +432 \eta ^2\right)\nonumber\\
&&+(1+2 \eta )^2 \left(9+150 \eta +880 \eta ^2+2080 \eta ^3+1536 \eta ^4\right)\nonumber\\
&&+2 J^2 (1+6 \eta ) \left(2 \alpha ^4+(1+2 \eta )^2 \left(5+16 \eta +8 \eta ^2\right)+\alpha ^2 \left(5+18 \eta +16 \eta ^2\right)
\right)\\
Da&=&6 \alpha ^6+\alpha ^4 \left(9+66 \eta +144 \eta ^2+J^2 (1+6 \eta )\right)+3 \eta ^2 (1+2 \eta )\nonumber\\
&&\times \left(9+J^4+96 \eta +304 \eta ^2+256 \eta ^3+2 J^2 \left(5+16 \eta +8 \eta ^2\right)\right)\nonumber\\
&&+2 \alpha ^2 \eta  \left(J^2 \left(7+23 \eta +24 \eta ^2\right)+3 \left(9+63 \eta +152 \eta ^2+144 \eta ^3\right)\right)\\
Db_1&=&\alpha  \left(6 \alpha ^4+\alpha ^2 \left(9+48 \eta +120 \eta ^2+J^2 (1+6 \eta )\right)+\eta  (1+2 \eta ) \left(J^2 (11+12 \eta )+3 \left(9+60 \eta +64 \eta ^2\right)\right)\right)\\
Db_2&=&J \alpha  \eta  \left(3-10 \alpha ^2+22 \eta +32 \eta ^2+J^2 (3+6 \eta )\right)\\
c_1&=&b_1\\
c_2&=&b_2\\
Dd_1&=&\alpha ^4 (6+8 \eta )+\alpha ^2 \left(9+48 \eta +120 \eta ^2+160 \eta ^3+J^2 \left(1+4 \eta +8 \eta ^2\right)\right)\nonumber\\
&&+\eta  (1+2 \eta ) \left(9+J^4+96 \eta +304 \eta ^2+256 \eta ^3+2 J^2 \left(5+16 \eta +8 \eta ^2\right)\right)\\
Dd_2&=&-J \alpha ^2 \left(9+6 \alpha ^2+66 \eta +96 \eta ^2+J^2 (1+6 \eta )\right)\\
De&=&6 \alpha ^6+\alpha ^4 \left(15+74 \eta +144 \eta ^2+J^2 (1+6 \eta )\right)+\eta  \left(1+5 \eta +6 \eta ^2\right) \nonumber\\
&&\times\left(9+J^4+96 \eta +304 \eta ^2+256 \eta ^3+2 J^2 \left(5+16 \eta +8 \eta ^2\right)\right)\nonumber\\
&&+\alpha ^2 \left(9+102 \eta +498 \eta ^2+1072 \eta ^3+864 \eta ^4+J^2 \left(1+18 \eta +54 \eta ^2+48 \eta ^3\right)\right)\\
f_1&=&d_1\\
f_2&=&0\\
Dg_1&=&\alpha  \left(9+6 \alpha ^4+105 \eta +418 \eta ^2+680 \eta ^3+384 \eta ^4+5 \alpha ^2 \left(3+16 \eta +24 \eta ^2\right)\right)\nonumber\\
&&+\alpha J^2 \left(1+13 \eta +34 \eta ^2+24 \eta ^3+\alpha ^2 (1+6 \eta )\right)\\
Dg_2&=&-J \alpha  \left(9+81 \eta +206 \eta ^2+160 \eta ^3+\alpha ^2 (6+22 \eta )+J^2 \left(1+5 \eta +6 \eta ^2\right)\right)\\
h&=&e\\
i_1&=&g_1\\
i_2&=&g_2
\end{eqnarray}

\section{Solution of Eq.(\ref{leq2})}

\begin{eqnarray}
D&=&J^4 (1+2 \eta )^3 (1+4 \eta )+(3+8 \eta ) \left(1+2 \alpha ^2+6 \eta +8 \eta ^2\right)^2 \left(3+2 \alpha ^2+10 \eta +8 \eta ^2\right)\nonumber\\
&&+2 J^2 (1+2 \eta ) \left(\alpha ^4 (2+8 \eta )+(1+2 \eta )^2 \left(5+32 \eta +56 \eta ^2+32 \eta ^3\right)+\alpha ^2 \left(5+38 \eta +96 \eta ^2+64 \eta ^3\right)\right)\\
Da&=&2 \alpha ^6 (3+8 \eta )+\alpha ^4 \left(9+66 \eta +184 \eta ^2+192 \eta ^3+J^2 \left(1+6 \eta +8 \eta ^2\right)\right)\nonumber\\
&&+\eta ^2 \left(1+6 \eta +8 \eta ^2\right) \left(9+J^4+72 \eta +176 \eta ^2+128 \eta ^3+2 J^2 \left(5+12 \eta +8 \eta ^2\right)\right)
\nonumber\\
&&+2 \alpha ^2 \eta  \left(9+93 \eta +352 \eta ^2+592 \eta ^3+384 \eta ^4+J^2 \left(3+23 \eta +48 \eta ^2+32 \eta ^3\right)\right)\\
Db_1&=&\alpha  \left(2 \alpha ^4 (3+8 \eta )+\eta  \left(1+6 \eta +8 \eta ^2\right) \left(9+36 \eta +32 \eta ^2+J^2 (5+4 \eta )\right)\right)\nonumber\\
&&+\alpha ^3 \left(9+60 \eta +144 \eta ^2+128 \eta ^3+J^2 \left(1+6 \eta +8 \eta ^2\right)\right)\\
Db_2&=&J \alpha  \eta  \left(J^2 \left(1+6 \eta +8 \eta ^2\right)-(3+8 \eta ) \left(1+2 \alpha ^2+6 \eta +8 \eta ^2\right)\right)\\
c_1&=&b_1\\
c_2&=&b_2\\
Dd_1&=&\alpha ^2 \left(J^2 (1+2 \eta )+(3+8 \eta ) \left(3+2 \alpha ^2+10 \eta +8 \eta ^2\right)\right)\\
Dd_2&=&-\left(J \alpha ^2 \left(J^2 \left(1+6 \eta +8 \eta ^2\right)+(3+8 \eta ) \left(3+2 \alpha ^2+14 \eta +8 \eta ^2\right)\right)\right)
\\
De&=&2 \alpha ^6 (3+8 \eta )+\alpha ^4 (1+2 \eta ) \left(15+76 \eta +96 \eta ^2+J^2 (1+4 \eta )\right)\nonumber\\
&&+\eta  \left(1+7 \eta +14 \eta ^2+8 \eta ^3\right) \left(9+J^4+72 \eta +176 \eta ^2+128 \eta ^3+2 J^2 \left(5+12 \eta +8 \eta ^2\right)\right)\nonumber\\
&&+\alpha ^2 \left(9+114 \eta +570 \eta ^2+1408 \eta ^3+1696 \eta ^4+768 \eta ^5+J^2 \left(1+18 \eta +78 \eta ^2+128 \eta ^3+64 \eta ^4\right)\right)\\
f_1&=&d_1\\
f_2&=&0\\
Dg_1&=&\alpha  \left(1+\alpha ^2+5 \eta +4 \eta ^2\right) \left(J^2 \left(1+6 \eta +8 \eta ^2\right)+(3+8 \eta ) \left(3+2 \alpha ^2+10 \eta +8 \eta ^2\right)\right)\\
Dg_2&=&-\left(J \alpha  \left(J^2 \left(1+7 \eta +14 \eta ^2+8 \eta ^3\right)+(3+8 \eta ) \left(\alpha ^2 (2+6 \eta )+3 \left(1+7 \eta +14 \eta ^2+8 \eta ^3\right)\right)\right)\right)\\
h&=&e\\
i_1&=&g_1\\
i_2&=&g_2
\end{eqnarray}

\end{widetext}


\acknowledgments
SM acknowledge the financial support of the European Commission, under the FET-Open grant agreement HIP, number FP7-ICT-221889.


\begin{thebibliography}{99}

\bibitem{vedral07}
V. Vedral, \emph{Introduction to Quantum Information Science},
Oxford University Press, Oxford (2007).

\bibitem{braun02}
D. Braun, Phys. Rev. Lett. \textbf{89}, 277901 (2002);
F. Benatti, R. Floreanini and M. Piani,
Phys. Rev. Lett. \textbf{91}, 070402 (2003).

\bibitem{PH02}
M. B. Plenio and S. F. Huelga, Phys. Rev. Lett. \textbf{88}, 197901 (2002)

\bibitem{clark03}
S. Clark, A. Peng, M. Gu and S. Parkins , Phys. Rev. Lett. \textbf{91}, 177901 (2003);
S. Mancini and J. Wang, Eur. Phys. J. D \textbf{32}, 257 (2005);
L. Hartmann, W. Dur, H.J. Briegel, Phys. Rev. A \textbf{74}, 052304 (2006);
D. Angelakis, S. Bose and S. Mancini, Europhys. Lett. \textbf{85}, 20007 (2007);S.F. Huelga, A. Rivas, and M.B. Plenio, arXiv:1106.2841v1.

\bibitem{mem11}
L. Memarzadeh and S. Mancini, Phys. Rev. A \textbf{83}, 042329 (2011).

\bibitem{polzik10}
H. Krauter, C. A. Muschik, K. Jensen, W. Wasilewski, J. M. Petersen, J. I. Cirac, E. S. Polzik,
arXiv:1006.4344.

\bibitem{march}
L. Gammaitoni, P. H\"anggi, P. Jung and F. Marchesoni, Rev. Mod. Phys. \textbf{70}, 223 (1998).


\bibitem{dots}
M. Trif, V. N. Golovach, D. Loss 	Phys. Rev. B \textbf 75, 085307 (2007))

\bibitem{ions}
A. Friedenauer, H. Schmitz, J. T. Glueckert, D. Porras and T. Schaetz, Nat. Phys. \textbf {4}, 757 (2008).

\bibitem{coldatoms}
S. Trotzky, P. Cheinet, S. Folling, M. Feld, U. Schnorrberger, A. M. Rey, A. Polkovnikov, E. A. Demler, M. D. Lukin, and I. Bloch, \textbf {319} 293 (2008).

\bibitem{CCAs}
A. Kay and D.G. Angelakis Eur. Phys. Lett. \textbf{ 84}, 20001 (2008).

\bibitem{Woo}
W.K. Wootters, Phys. Rev. Lett. \textbf{80}, 2245 (1998).

\bibitem{ALK10}
D. Angelakis, D. Li and LC Kwek, Europhys. Lett. \textbf{91}, 10003 (2010).

\bibitem{GK76}
V. Gorini and A. Kossakowski, J. Math. Phys. \textbf{17}, 1298  (1976).



\bibitem{huelga09}
A. Rivas, N. P. Oxtoby and S. F. Huelga, Eur. Phys. J. B \textbf{69}, 51 (2009).





\end{thebibliography}
\end{document}